\documentclass{elsart}

\usepackage{amsfonts}
\journal{Annals of Physics}

\DeclareFontFamily{U}{msa}{}
\DeclareFontShape{U}{msa}{m}{n}
    { <-> msam10}{}
\DeclareSymbolFont{AMSa}{U}{msa}{m}{n}

\DeclareFontFamily{U}{msb}{}
\DeclareFontShape{U}{msb}{m}{n}
     { <-> msbm10}{}
\DeclareSymbolFont{AMSb}{U}{msb}{m}{n}

\DeclareFontFamily{U}{euf}{}
\DeclareFontShape{U}{euf}{m}{n}
    { <-> eufm10}{}
\DeclareFontShape{U}{euf}{b}{n}
    { <-> eufb10}{}

\DeclareFontFamily{U}{eus}{}
\DeclareFontShape{U}{eus}{m}{n}
    { <-> eusm10}{}
\DeclareFontShape{U}{eus}{b}{n}
    { <-> eusb10}{}

\DeclareFontFamily{U}{eur}{}
\DeclareFontShape{U}{eur}{m}{n}
    { <-> eurm10}{}
\DeclareFontShape{U}{eur}{b}{n}
    { <-> eurb10}{}

\DeclareMathAlphabet{\matheurm}{U}{eur}{m}{n}
\DeclareMathAlphabet{\matheuf}{U}{euf}{m}{n}
\DeclareMathAlphabet{\matheurmbf}{U}{eur}{b}{n}
\DeclareMathAlphabet{\matheuscr}{U}{eus}{m}{n}
\DeclareMathAlphabet{\mathsfsl}{OT1}{cmss}{m}{sl}
\DeclareMathAlphabet{\mathsf}{OT1}{cmss}{m}{n}
\DeclareFontShape{OT1}{cmr}{bx}{n}{ <-> cmbx10 }{}

\DeclareMathSymbol{\smallfrown}{\mathrel}{AMSa}{"61}
\DeclareMathSymbol{\subsetneq}{\mathrel}{AMSb}{"24}
\DeclareMathSymbol{\therefore}{\mathrel}{AMSa}{"29}
\DeclareMathSymbol\compl{\mathord}{AMSb}{"73}
\DeclareMathSymbol\restriction{\mathord}{AMSa}{"18}

\newcommand{\disptitle}[1]{{\sc #1}}

\newcommand{\calcmd}[1]{{\mathcal #1}}

\newcommand{\concat}{{}^\smallfrown}
\newcommand{\geom}[1]{{\calcmd{#1}}}

\newcommand{\preset}[2]{{}^{#1}{}#2}

\newcommand{\vecsp}[1]{{\mathsf #1}}
\newcommand{\eqdef}{{\,\stackrel{\rm{def}}=}\,}

\newcommand{\image}{{}^{\scriptscriptstyle\rightarrow}}
\newcommand{\invimage}{{}^{\scriptscriptstyle\leftarrow}}

\newcommand{\elsalg}[1]{\mathfrak #1}
\newcommand{\system}[1]{\mathfrak{#1}}

\newcommand{\im}{\operatorname{im}}
\newcommand{\dom}{\operatorname{dom}}
\newcommand{\lspan}{\operatorname{span}}

\newtheorem{definition}{Definition}

\newtheorem{statement}{}

\renewcommand{\disptitle}[1]{{\sc #1}}

\newcommand{\parper}{\square}

\newcommand{\complet}[1]{{\overline{#1}}}
\newcommand{\setalg}{{\elsalg S}}

\newenvironment{elscases}{\left\{\begin{array}{ll}}{\end{array}\right.}
\newcommand{\boldsymbol}[1]{{\bf{#1}}}
\newcommand{\implies}{\Longrightarrow}
\newcommand{\eg}{e.g.}
\newcommand{\ie}{i.e.}
\newcommand{\viz}{viz.}

\begin{document}
\section{Introduction}

The quantum and classical worldviews differ
profoundly in two ways, both of which bear on the nature of
observation.  Briefly, if $\system A$ and
$\system B$ are quantum systems whose states are represented in
Hilbert spaces $\vecsp A$ and $\vecsp B$ then the
states of the composite system $\system {AB}$ are represented in
the tensor product $\vecsp A \otimes \vecsp B$.  Only if the state
of $\system{AB}$ is a pure product state can
$\system A$ and $\system B$ be truly regarded as separate entities.  If
$\system A$ and $\system B$ interact then in
general a pure product state will evolve to a mixed state.  This is
in sharp contrast to the classical worldview, according to which
$\system A$ and $\system B$ always have definite states, whether
they interact or not. 

This entanglement of interacting systems becomes particularly
interesting when the interaction has the nature of
\emph{observation}, and it is closely related
to the other great novelty of the quantum theory, viz., the
apparent stochastic nature of measurement in the quantum
theory.  We say `apparent' because the conclusion
that quantum observation is nondeterministic follows from the
insistence that an observed system $\system S$ and an observer $\system O$ are in definite states following an observation.

Suppose $\system S$ and $\system O$ are represented in 
Hilbert spaces $\vecsp U$ and $\vecsp W$.  Suppose for convenience that the measurement in question corresponds to a selfadjoint operator $A$ on $\vecsp U$ with a discrete spectrum $\{ \lambda_i \}$. 
Let $\vecsp U_i$ be the $\lambda_i$-eigenspace of
$A$, and let $P_i$ be the orthogonal projection (OP) to $\vecsp U_i$. 
Then $\{ \vecsp U_i \}$ is an orthogonal decomposition
of $\vecsp U$, and $\{ P_i \}$ is an orthogonal
decomposition of the identity.  In order that the
interaction of $\system S$ and $\system O$ qualify as a measurement
of $A$, final states of $\system O$ corresponding to distinct
eigenvalues of $A$ must be distinguishable with certainty, \ie,
they must be mutually orthogonal.  Let $\vecsp W_i$ be the subspace
of $\vecsp W$ corresponding to the value $\lambda_i$.  Suppose
$\theta \in \vecsp U$ and $\varphi \in \vecsp W$ are unit vectors. 
A measurement interaction is defined as one in which an initial
state $\theta \otimes \varphi$ evolves to a state
\begin{equation}
\label{cba}
\sum_i P_i \theta \otimes \varphi_i
 =
\sum_i \theta_i \otimes \varphi_i,
\end{equation}
where, for each $i$, $\varphi_i$ is a unit vector in $\vecsp
W_i$, and we have defined $\theta_i = P_i \theta$ for notational
convenience.

The Copenhagen interpretation states that if $\system O$ is an
observer in some sense, \eg, if $\system O$ is a person reading an
experimental apparatus, then the final state of $\system {SO}$ is
not the sum (\ref{cba}) but is actually one of the pure product
states $\theta_i \otimes \varphi_i$, and the probability that it is
$\theta_i \otimes \varphi_i$ is $\| \theta_i \otimes \varphi_i \|^2
= \| \theta_i \|^2$.  In this way the Copenhagen interpretation accounts
for our subjective experience that when we perform a measurement as
above we obtain a definite value; and it also accounts for our
empirical observation that if the measurement is repeated many
times, the frequency with which each value $\lambda_i$ is observed
is close to $\| \theta_i \|^2$.  Note that the interpretation of
$\| \theta_i \|^2$ as a probability is ``written into'' the
Copenhagen interpretation, which is to say, it is taken as an axiom
of the quantum theory.  Following Farhi \emph{et al}\cite{Farhi:1989}, we refer to this as the \emph{probability postulate}.

The alternative to the Copenhagen interpretation is to take
(\ref{cba}) at face value, as the actual final state of $\system
{SO}$.  This is really no interpretation at all---being just a restatement \emph{with emphasis} of the basic principles of the quantum theory of interacting systems---and hardly
deserves a name, but its metaphysical implications are felt by
some to be so counterintuitive that it must be---if not rejected
outright---at least subjected to critical examination, for which
purpose it has been given the evocative name of the
\emph{many-worlds interpretation}.

The Copenhagen interpretation is useful in that it allows one to
``get on'' with the business of physics insofar as this is
understood to mean making and confirming theoretical predictions of
the results of experiments.  As a description of reality, however,
it is defective, principally in that there is no objective
criterion to decide when an interaction of physical systems
proceeds according to the ordinary laws of quantum physics and when
it undergoes the so-called \emph{wave-packet collapse} described above.  A vivid version of this conundrum is Schr\"odinger's cat, who
is either dead or alive inside a box depending on whether a
radioactive nucleus has decayed, the question being whether
the animal is really either one or the other before some higher
form of life (read: the scientist performing this callous
experiment) opens the box to observe the result.

The manifest impossibility of defining which interactions are
observations in the Copenhagen sense (quite apart from the evident
impossibility of saying anything intelligible about the actual
process of wave-packet collapse) would appear to rule out the
Copenhagen interpretation as a model of reality.  On the other
hand, the principal (indeed, upon close analysis the only)
objection to the many-worlds ``interpretation'' seems to be:  ``I
don't know about you, but \emph{I}'m certainly in a
definite state''.  This sense that the many-worlds
interpretation of (\ref{cba}) is incompatible with experience
appears ill founded on closer examination, however, inasmuch as any
question posed of $\system {SO}$---in the form of a selfadjoint operator to be measured---that has the answer `yes' with certainty for each
$\theta_i \otimes \varphi_i$, has the answer `yes' with certainty
for $\sum_i \theta_i \otimes \varphi_i$, and \emph{vice versa}. 
(If we let `yes' correspond to the eigenvalue 1, then such an
operator is the identity.)  In other words, one cannot effectively
ask the question:  `Is the world in a pure product state or in a
superposition of such states?'.

As far as the analysis of a single measurement is concerned, the
many-worlds interpretation thus has no disadvantage, and since it
has the advantage of internal consistency it would seem to be the
interpretation of choice.  Note, however, that in the many-worlds
view $\| \theta_i \|^2$ is merely the weight of $\theta_i
\otimes \varphi_i$ in $\sum_i \theta_i \otimes \varphi_i$, rather
than the probability of $\theta_i \otimes \varphi_i$, as in the
Copenhagen interpretation.  Indeed, the many-worlds interpretation
of a single measurement makes no statement at all regarding
probability.

The interpretation of $\| \theta_i \|^2$ as a probability is of course based on the fact that in a series of repetitions of this
measurement the cumulative frequency of outcome $\lambda_i$
approaches $\| \theta_i \|^2$.  This raises the following
intriguing issue.  In the many-worlds interpretation all possible
outcomes of a measurement operation are represented, and in the
case of repeated measurements, all possible sequences of outcomes
are represented.  Does this not imply the certainty that
whenever we carry out an experiment of this sort there are versions
of us who witness sequences of outcomes wildly at variance with
theoretical predictions?  And are these versions not
therefore led to reject the quantum theory solely
because of bad luck?  The answer of course is `yes', just as in the
Copenhagen interpretation the possibility is always present that relative frequencies of outcomes in a series of measurements will deviate significantly from theoretical
prediction.  Note that in the Copenhagen interpretation this is
only a possibility, while in the many-worlds interpretation it is
a certainty.  The probability is small in the Copenhagen
interpretation, just as the magnitude of the corresponding
component is small in the many-worlds interpretation.  The
difference is that while we have learned to feel comfortable
ignoring possible outcomes that have small probability (indeed, our
sanity depends on it), we do not have a rationale for ignoring
components with small magnitudes---we cannot say that we are
unlikely to have the corresponding experience, because some version
of us certainly \emph{does} have that experience. 

Our instinct when confronted with what we suspect to be a
statistical deviation is to prolong the series of measurements
until the observed frequencies come back into line.  In the
Copenhagen interpretation, by the strong law of large numbers, this
happens with probability 1.  Precisely, in an infinite series of
measurements, the probability that the frequency statistic will not
tend to its proper limit is 0.  It is natural to ask whether this
scenario may be modeled in the many-worlds interpretation and
whether the component corresponding to improper limiting behavior
of the frequency statistic may be shown to vanish.  In fact, this would appear to be the \emph{essential} question, a positive answer to which nullifies any remaining objection to the many-worlds view.

Everett\cite{Everett:1957} discussed the interpretation of $\| \theta_i \|^2$ as a probability in the many-worlds interpretation, and Hartle\cite{Hartle:1968} and Graham\cite{Graham:1973} applied the weak law of large numbers to the frequency statistic in the context of finitely many observations; but to actually derive the probability interpretation of the squared norm from prior principles one must apply the strong law of large numbers to infinite sequences of observables.

The simplest generalization of (\ref{cba}) to $N$
independent measurements involves the tensor product of $N+1$
statevector spaces, and the corresponding generalization to
infinitely many measurements involves the tensor product of
infinitely many Hilbert spaces.  As shown by von Neumann~\cite{vonNeumann:1938}, such a product can be defined as
a complex inner product space, but it is not separable---\ie, it has an uncountable set of pairwise orthonormal vectors.  It thus
lies outside the standard framework of the quantum theory.  

An analysis of the many-worlds interpretation for infinite sequences of observations in terms the von Neumann product was carried out by Farhi, Goldstone, and Gutmann\cite{Farhi:1989} (see also \cite{Gutmann:1995}).  Unaware of the von Neumann construction, and \emph{a fortiori} unaware of the work of Farhi \emph{et al.}, the present author conceived and derived this result in the context of ordinary (separable) Hilbert space.  Here (perceived) necessity was indeed the mother of invention, as the construction employed in this proof was the inspiration and starting point for the construction of the hidden-variables models presented in \cite{RVWhv:2005}---in which, as in the many-worlds setting, the apparent stochastic nature of observation is inherent in the model, which is not itself probabilistic.  \cite{RVWhv:2005} demonstrates a natural correspondence between hidden-variables states and \emph{generic} objects of the sort that figure prominently in the foundations of mathematics---a correspondence with intriguing ontological implications for hidden-variables states.

As explained above, the novelty of the present analysis compared to \cite{Farhi:1989,Gutmann:1995} is that it takes place entirely within the conventional framework of quantum mechanics, in which physical states are represented by vectors in ordinary Hilbert space, as opposed to a \emph{nonseparable} inner-product space in which there exists an uncountable set of pairwise orthogonal vectors.  Cassinello and S{\'a}nchez-G{\'o}mez\cite{Cassinello:1996} and Caves and Schack\cite{Caves:2005} have criticized \cite{Farhi:1989}, claiming that it does not succeed in its goal of deriving the probability postulate from prior quantum principles.  We disagree with these authors and present a critique of their critique in Section~\ref{sec crit crit}.

As a practical matter, it will be convenient to employ certain elementary conventions of modern set theory throughout this discussion; these conventions are summarized in the appendix along with the definitions and basic properties of partial orders and boolean algebras.  Theorems of a general nature are labeled `{\bf Proposition}'; proofs of these are straightforward or readily available in the existing literature, but some are nevertheless presented here for their pedagogic value.  Theorems specific to the argument of this paper are labeled `{\bf Theorem}'.

\section{Propositional algebras}
\label{sec prop alg}

Suppose $\vecsp V$ is a Hilbert space.  We use the term \emph{proposition} to refer to an orthogonal projection or to its 1-eigenspace, \ie, its image, according to context.  Any closed subspace $\vecsp P$ of $\vecsp V$ is the image of a unique orthogonal projection, the \emph{orthogonal projection to $\vecsp P$}.  We say that propositions $P$ and $Q$ are
\emph{orthogonal} to one another iff they are orthogonal as spaces, \ie, all vectors in $P$ are orthogonal to all vectors in $Q$.  Regarding propositions as projections, this is equivalent to $\im P \perp \im Q$, and this is equivalent to either of the conditions: $\im Q
\subseteq \ker P$ or $\im P \subseteq \ker Q$, and therefore to
either of the conditions: $PQ = 0$ or $QP = 0$ (`0' here denoting the zero operator).

We define the \emph{complement} $\compl P$ of a proposition $P$ to be $P^\perp \eqdef \{ \phi \in \vecsp V \mid (\forall \psi \in P) \phi \perp \psi \}$, regarding $P$ as a space, or equivalently, $1-P$, regarding $P$ as a projection.  We say that propositions $P$ and $Q$ \emph{commute}, $P \parper Q$, iff they commute as projections.  If $P \parper Q$ we define the \emph{meet} $P \wedge Q$ and \emph{join} $P \vee Q$ by
\[
P\wedge Q = P \cap Q \mbox{\ and\ } P \vee Q = \lspan (P \cup Q),
\]
treating $P$ and $Q$ as spaces, where $\lspan$ is the linear span, \ie, closure under vector addition and scalar multiplication.  Note that commuting propositions correspond to \emph{perpendicular} spaces, spaces $\vecsp P$ and $\vecsp Q$ being perpendicular iff there exist pairwise orthogonal spaces $\vecsp A$, $\vecsp B$, and $\vecsp C$ such that
\[
\vecsp P \cap \vecsp Q = \vecsp A,\ \vecsp P = \vecsp A \oplus \vecsp B,\ \mbox{and}\ \vecsp Q = \vecsp A \oplus \vecsp C.
\]
$\oplus$ is \emph{direct sum}, \ie, the linear span of the union of two spaces whose intersection is $\{0\}$ (`0' here denoting the zero vector).

Suppose $\elsalg A$ is a set of commuting propositions closed under the operations of
complementation and meet (equivalently, join).  Then $\elsalg A$ is a \emph{boolean
algebra} (see Def.~\ref{def bool alg} in the appendix), specifically a \emph{propositional algebra} (PA).

Note that if $\mathcal A$ is a set of
commuting propositions and $P, Q \in \mathcal A$, then
$1-P$ and $PQ$ commute with all members of $\mathcal A$.  Thus any
set $\mathcal A$ of commuting propositions may be 
closed under complementation and meet to form the boolean
algebra \emph{generated by $\mathcal A$}, which is the smallest
propositional algebra that includes
$\mathcal A$.

The meet $\bigwedge \mathcal B$ of a set $\mathcal B$
of commuting propositions is $\bigcap \{ P \mid P \in
\mathcal B \}$, which, as the intersection of closed spaces, is closed and is therefore a proposition.  Note that if $\mathcal A$ is a set
of commuting propositions and $\mathcal B \subseteq
\mathcal A$, then $\bigwedge \mathcal B$ commutes with all members
of $\mathcal A$.  Thus any set $\mathcal A$ of commuting propositions may be may be closed under complementation and
arbitrary meets (or joins) to form the complete propositional
algebra \emph{generated by $\mathcal A$}, which is the smallest
complete PA that includes $\mathcal A$.

The paradigm of a boolean algebra is the collection $\setalg(A)$
of all subsets of a given set $A$.  In this algebra meet, join, and
complement are respectively intersection, union, and set-theoretic
complement relative to $A$ ($\compl B = A \setminus B$).  A subset
of $\setalg(A)$ is a boolean algebra iff it is closed under
complementation and intersection (and hence also union).  We will
be interested in countably complete subset algebras, which are
closed under countable intersections (and countable unions).

For the present discussion we are interested in the case $A =
\preset{\omega}2$, the set of functions from $\omega$ to $2$, where $\omega = \{ 0, 1, \dots \}$ and $2 = \{
0, 1 \}$.\footnote{The use of `2' to denote $\{0,1\}$ is standard in set theory when 0, 1, and 2 are ordinals.  For certain purposes it is convenient to identify the ordinals 0 and 1 with the real numbers with the same names, but in expressions like `$\preset\omega2$', `2' denotes the set $\{0,1\}$, not the real number 2.}  For $n \in \omega$ and $\epsilon \in 2$, let
$J^n_\epsilon = \{ f \in \preset{\omega}2 \mid f(n) = \epsilon \}$. 
The \emph{standard topology} on $\preset{\omega}2$ is defined by
taking $\{ J^n_\epsilon \mid n \in \omega,\ \epsilon \in 2 \}$ as
a subbase.  That is, the open sets are arbitrary unions of finite
intersections of $J^n_\epsilon$s.  The closed sets are those whose
complements are open.  The class $\elsalg B$ of Borel sets is the
closure of the class of open (or closed) sets under the operations
of complementation and countable intersection.  $\elsalg B$ is clearly
a countably complete boolean algebra.

A map $h : \elsalg A \to \elsalg C$ from one boolean algebra
into another is a \emph{homomorphism} iff it commutes with the complement
operation and the binary meet operation (or, equivalently, the
binary join operation).  $h$ is \emph{countably
complete} or \emph{complete} iff it also commutes with countable meets and joins or arbitrary meets and joins, respectively. 
\begin{thm}
\label{cas}
Suppose $P_n$, $n \in \omega$, are commuting propositions.  There exists a countably complete homomorphism $X
\mapsto P(X)$ from the boolean algebra $\elsalg B$ of Borel subsets of
$\preset{\omega}2$ to the complete PA generated by $\{ P_n \mid n \in \omega \}$ such that
for each $n \in \omega$,
\begin{equation}
\label{cat}
P(J^n_\epsilon) =
 \begin{elscases}
  P_n & \mbox{if $\epsilon = 1$}\\
  1 - P_n & \mbox{if $\epsilon = 0$}.
 \end{elscases}
\end{equation}
\end{thm}
\begin{pf} To extend (\ref{cat}) to a homomorphism on $\elsalg B$ we could proceed by repeatedly adjoining complements and
countable intersections, defining $P(\preset{\omega}2 \setminus X)
= \compl P(X)$ and $P\left(\bigcap_{n \in \omega} X_n
\right) = \bigwedge_{n \in \omega} P(X_n)$.  To effect this
construction we would have to show that at each intermediate stage,
if we have families $\{ X_n \mid n \in \omega \}$ and $\{ Y_n \mid
n \in \omega \}$ such that $P(X_n)$ and $P(Y_n)$ are defined for
all $n \in \omega$, and $\bigcap_{n \in \omega} X_n = \bigcap_{n
\in \omega} Y_n$, then $\bigwedge_{n \in \omega} P(X_n) =
\bigwedge_{n \in \omega} P(Y_n)$.  This can be done directly, but
it is easier to use the spectral theory of selfadjoint operators
on Hilbert space.\footnote{We are admittedly using an elephant gun to slay a gnat, but it gets the job done.}
\begin{definition}
\label{czo}
For $n \in \omega$ and $\sigma \in \preset{n}2$, \ie, for $\sigma$
a sequence of 0s and 1s of length $n$, define $I_\sigma =
\bigcap_{m < n} J^m_{\sigma(m)}$.  Thus, $I_\sigma = \{ f \in
\preset{\omega}2 \mid \sigma \subseteq f \}$.\footnote{Remember that functions are sets of ordered pairs, so $\sigma \subseteq f$ iff the finite sequence $\sigma$ is an initial segment of the infinite sequence $f$.}
\end{definition}
Let
\begin{equation}
\label{eq P I sigma}P(I_\sigma) = \prod_{m < n} P(J^m_{\sigma(m)}).
\end{equation}
Define a map $\rho : \preset{\omega}2 \to [0,3/2]$ by
\begin{equation}
\label{eam}
\rho(f) = \sum_{n=0}^\infty f(n) 3^{-n}.
\end{equation}
Note that for $\sigma \in \preset{n}2$, $\rho\image I_\sigma
\subseteq C_\sigma \eqdef [x_\sigma, x_\sigma + 3^{-n}(3/2)]$,
where
\[
x_\sigma = \sum_{m=0}^{n-1} \sigma(m) 3^{-m}.
\]
$\rho$ is a homeomorphism of $\preset{\omega}2$ with the Cantor set
formed from $[0,3/2]$.  (The Cantor set of a closed interval is
obtained by deleting the middle open third of the interval, then
deleting the middle open third of each of the two remaining closed
intervals, then deleting the middle open thirds of each of the four
remainders, \emph{ad infinitum}, and taking the intersection of all
of the closed sets of this sequence.)  Let $P_\lambda = \bigvee
\{P(I_\sigma) \mid \sigma \in \preset{{<\omega}}2\ \&\ \rho\image
I_{\sigma} \subseteq [0,\lambda] \}$.  Let $A$ be the selfadjoint
operator defined by\footnote{Eq. (\ref{eq spectral thm}) means that for any $\psi \in \vecsp V$, $A\psi$ is the limit of sums of the form $\sum_{i=1}^n \lambda_i (P_{\lambda_i} - P_{\lambda_{i-1}}) \psi$, with $\lambda_0 < \lambda_1 < \cdots < \lambda_n$, as $\lambda_0 \to -\infty$, $\lambda_n \to \infty$, and $\max_{i=1, \dots,n} (\lambda_i - \lambda_{i-1}) \to 0$, if the limit exists and is otherwise undefined.  The spectral theorem states that the selfadjoint operators are exactly the operators of this form for monotone increasing OP-valued functions $\lambda \mapsto P_\lambda$ for which $\lim_{\lambda \to -\infty} P_\lambda = 0$ and $\lim_{\lambda \to \infty} P_\lambda = 1$.  Such functions are called \emph{projection-valued measures}.}
\begin{equation}
\label{eq spectral thm}
A \eqdef \int \lambda\, dP_\lambda.
\end{equation}
For any Borel function $F : [0,3/2] \to \mathbb R$, the integral
$\int F(\lambda)\, dP_\lambda$ is well defined, and this is the
standard definition of $F(A)$.  In particular, suppose $X \subseteq
\preset{\omega}2$ is a Borel set.  Let $F_X : [0,3/2] \to \mathbb
R$ be the characteristic function of $\rho\image X$, \ie,
\[
F_X(\lambda)
 =
\begin{elscases}
 1 & \mbox{if $\rho^{-1} (\lambda) \in X$}\\
 0 & \mbox{if $\rho^{-1} (\lambda) \notin X$}.
\end{elscases}
\]
Let
\[
P(X) = F_X(A).
\]
$P(X)$ is a selfadjoint operator with spectrum included in $\{ 0,
1\}$, so it is an orthogonal projection.  One easily shows that the
map $X \mapsto P(X)$ is a homomorphism of $\elsalg B$ with a PA such that (\ref{cat}) holds.  Clearly
the image of this map is the complete boolean algebra generated by
$\{ P_n \mid n \in \omega \}$.\qed\end{pf}

From now on we suppose that $\langle P_n \mid n \in \omega \rangle$ is an $\omega$-sequence
of commuting propositions, $\elsalg P$ is the complete
PA it generates, and $X \mapsto P(X)$ is the homomorphism of the Borel algebra $\elsalg B$
to $\elsalg P$ as given by Theorem~\ref{cas}.

Let $\psi \in \vecsp V$ be an arbitrary nonzero vector.  Define
\begin{equation}
\label{eaj}
\mu_\psi(X) = \| P(X) \psi \|^2.
\end{equation}
One readily verifies that $\mu_\psi$ is a measure on $\elsalg B$,
\ie,
\begin{enumerate}
\item $\mu_\psi(\emptyset) = 0$,
\item $\mu_\psi(X) \ge 0$ for all $X \in \elsalg B$, and
\item if $X = \bigcup_{n \in \omega} X_n$ and $X_m \cap X_n =
\emptyset$ for all $m \neq n$, then
\[
\mu_\psi(X) = \sum_{n=0}^\infty \mu_\psi(X_n).
\]
\end{enumerate}
Note that $\mu_\psi(\preset{\omega}2) = \| \psi \|^2$.

The foregoing situation has the following physical interpretation. 
A proposition corresponds to a measurement with two
possible outcomes:  0 and 1.  We may paraphrase
the measurement corresponding to $P$ as `what is the value of $P$?'
or `is $P$ 1?'.  The projections $P_n,\ n \in \omega$, correspond
to simultaneously answerable questions about a physical state.  For
any Borel set $X \subseteq \preset{\omega}2$, $P(X)$ corresponds to
the question `is the sequence of values of $\langle P_n \mid n \in
\omega \rangle$ an element of $X$?'.

\section{Probability}
\label{sec prob}

Suppose $\psi$ is a normalized statevector and $0 < q < 1$.  We say that $\psi$ is
\emph{$q$-homogeneous} for $\langle P_n \mid n \in \omega \rangle$
iff for any $n \in \omega$ and any $\sigma \in
\preset{n}2$,
\begin{equation}
\label{czq}
\|P(I_{\sigma\concat\langle 1 \rangle}) \psi\|
 =
q \|P(I_\sigma) \psi\|,
\end{equation}
If $\psi$ is $q$-homogeneous then $\mu_\psi$ is the \emph{$q$-homogeneous} measure $\mu_q$, which is the measure on $\preset{\omega}2$ defined by the condition that
\[
\mu_q(I_\sigma)
 =
\prod_{i=0}^{n-1} q_{\sigma(i)},
\]
for any $n \in \omega$ and $\sigma \in \preset n2$, where
\begin{eqnarray}
q_0 &=& 1-q\\
q_1 &=& q.
\end{eqnarray}
For example, $\psi$ could represent the Heisenberg state of an electron that is subject to alternating measurements of its spin along two perpendicular axes, $P_n$ representing the $n^{\mathrm{th}}$ observation, in which case $q = 1/2$.  For another example, if we imagine a system with infinite spatial extent, the $P_n$s could represent judiciously chosen simultaneous measurements at different places.  For a third example---which does not involve infinite extension in time or space---let $\psi$ be the state of a point particle uniformly distributed in a box $B$ that extends along the $x$-axis from 0 to 1.  For $n = 0, 1, \dots$, and $i = 0, 1$, let $B^n_i$ be the portion of $B$ consisting of points with $x$-coordinates whose binary expansion contains $i$ in the $n^{\mathrm{th}}$ place.  Let $P_n$ consist of determining whether the particle is in $B^n_1$.  The $P_n$s commute---indeed, they are all functions of the position operator for the $x$-coordinate---and $\psi$ is $\textstyle\frac12$-homogeneous for this sequence.  Each of these examples is infinitary in its own way---the last in that it contemplates measurements of arbitarily high precision, but this is intrinsic to the notion of a continuous observable and is standard in physics; indeed, the simultaneous performance of all the $P_n$s in this case is simply the measurement of $x$.  The impossibility of actually performing infinitely many---or infinitely precise---observations in any effective sense should not be taken to invalidate this line of reasoning, any more than it invalidates applied probability theory, which in the final analysis is founded on the notion of an infinite sequence of trials.

Let $L_q \subseteq \preset{\omega}2$ be defined by the condition
\begin{equation}
\label{ebe}
f \in L_q
 \iff
\lim_{N \to \infty} F(N) = q,
\end{equation}
where
\begin{equation}
\label{ebf}
F(N)
 \eqdef
\frac1N \sum_{n < N} f(n)
\end{equation}
is the \emph{frequency statistic}.
\begin{prop}
{\sc Strong law of large numbers}\[\mu_q(L_q) = 1.\]
\end{prop}
\begin{thm}
Thus, if $\psi$ is $q$-homogeneous then $\mu_{\psi}(L_q) = 1$, whence, by (\ref{eaj}), $P(L_q) \psi = \psi$, \ie,
$\psi$ is a 1-eigenvector of the observable $P(L_q)$, which
represents the question{\normalfont
\begin{statement}
\label{czr}
\mbox{`does the frequency $F(N)$ tend to $q$ as $N \to \infty$?'}.
\end{statement}}
Hence the answer to this question, posed of $\psi$, is `yes'.
\end{thm}

Note that $P(L_q)$ is in $\elsalg P$ and commutes with all $P \in \elsalg
P$.  Hence
\begin{equation}
\label{eq PPL}
P(L_q) P \psi = P P(L_q) \psi = 0
\end{equation}
for any $P \in \elsalg P$.  For $\psi \in \vecsp V$, let
\begin{equation}
\label{eak}
[\psi]^{\elsalg P}
 \eqdef
\overline{\lspan \{ P\psi \mid P \in \elsalg P \}},
\end{equation}
\ie, the closure of the linear span of the orbit of $\psi$ under
the action of $\elsalg P$.  $[\psi]^{\elsalg P}$ is a mode\footnote{A \emph{mode} of a physical system is the set of states corresponding to a closed subspace of its statevector space.} of the system
in question and is the smallest mode that contains $\psi$ and all
the states to which it may be projected as a result of observations
in $\elsalg P$.  (\ref{eq PPL}) gives
\begin{thm}
\label{eax}
$P(L_q)$ restricted to $[\psi]^{\elsalg P}$ is the identity, so the answer
to the question {\bf\ref{czr}} posed of any state in $[\psi]^{\elsalg P}$ is `yes'.
\end{thm}
This corresponds to the fact that the first few values in an infinite sequence of trials (where
`few' may refer to any finite number) have no bearing on its
behavior in the limit of infinitely many trials.

\section{Tests of randomness}
\label{ebk}

Note that there is nothing probabilistic in the statements of the preceding section.  We
are not saying that {\bf\ref{czr}} is usually true or even
almost always true,\footnote{`Almost always' is used in measure theory to refer to sets whose complements have measure 0, as, for example, the rational numbers have measure 0 as a subset of the real numbers with the usual Lebesgue measure.  In probability theory---which is just measure theory when the total measure is 1---`almost always' means `with probability 1', as, for example, a real number chosen at random according to a continuous probability distribution is almost always irrational.} but rather that it is simply true.  This is
what it means to be an eigenstate of an observable.  Isn't
this a bit strange?  Surely there is the possibility that $F(N)
\not\to q$ even though the probability is 0?  Not in the quantum
worldview, which is in this way more reasonable than the classical
worldview.  We will have more to say on this subject a little
later.

So where does probability come in?  Probability is just a handy way
of talking about positive measures with maximum value 1.  In the
preceding example, we may regard the sequences $f \in \preset\omega2$
as arising from the random process $\mathcal P_q$ that produces a sequence of 0s and 1s, with the probability of producing a 1 at any step being $q$.  For any $X \subseteq
\preset\omega2$, $\mu_q(X)$ is the probability that a sequence
generated by $\mathcal P$ will be in $X$.  $L_q$ may be regarded as
a \emph{test of `$\mathcal P = \mathcal P_q$'} in the following
sense.  Suppose we are presented with a black box that produces a
sequence of 0s and 1s according to some process $\mathcal P$.  If we used this device to generate a
sequence $f : \omega \to 2$ and found that $f \notin L_q$, we would
be justified in concluding that the device was not operating
according to the process $\mathcal P_q$, \ie, $\mathcal P \neq \mathcal P_q$.  Any Borel $X \subseteq \preset\omega2$ with $\mu_q(X) = 1$ will similarly serve as a test of the supposition that $\mathcal P = \mathcal P_q$, and $P(X)$ is definitely true of any state in $[\psi]^{\elsalg P}$.  This was first observed by Gutmann\cite{Gutmann:1995} in the nonseparable situation.

Note that any countable collection of such tests may be
applied simultaneously, because the intersection of any countable
collection of sets of measure 1 has measure 1.  Indeed we may
imagine a grand test of `$\mathcal P = \mathcal P_q$' that is the
intersection of every imaginable test of this supposition, since there
are only countably many things we can imagine (in truth only finitely many, but let's be generous).  Let $G_q \subseteq \preset{\omega}2$ be this set.\footnote{The use of imaginability as the criterion for inclusion in $G_q$ is admittedly somewhat facetious.  To be more objective we could replace it with definability \emph{in some sense}.  The objectivity of definability is something of an illusion, which is why we have emphasized `in some sense':  `define' has to be defined.  The potential difficulty is illustrated by Richard's paradox, which asks `what is the smallest natural number not definable by an English phrase of fewer than sixteen words?'.  Since there are only finitely many English phrases of fewer than sixteen words, there must be such a number, and there must therefore be a least such.  But this number is definable by the phrase
\[
\begin{array}{l}
\mbox{the smallest natural number not definable}\\
\mbox{\qquad by an English phrase of fewer than sixteen words},
\end{array}
\]
which has fifteen words:  a contradiction.  We return to this paradox in \cite[Section 6.1]{RVWhv:2005}, where we give its resolution as an important limitation on the definability of `define'.}

  Then
$\mu_q(G_q) = 1$, so $P(G_q) \psi = \psi$, and we see that the question
`does the sequence of outcomes of the observations $P_0, P_1,
\dots$, pass every imaginable test of `$\mathcal P = \mathcal P_q$'?', posed of $\psi$, has the definite answer `yes'.  Thus we may regard the process of measurement as stochastic.

The preceding example is only a special case of a general phenomenon.  Suppose $\boldsymbol P = \{ P_n \mid n \in \omega\}$ is an arbitrary family of commuting propositions, $\elsalg P$ is the PA it generates, and $\psi$ is an arbitrary statevector.  The probability measure $\mu^{\boldsymbol P}_\psi$ is in general not of the form $\mu_q$; rather, the probability that $P_n=1$ conditioned on the values of $P_m$ for one or more $m \neq n$, or more generally conditioned on the value of $Q$ for any $Q \in \complet{\elsalg P}$, is generally dependent on those values.  Nevertheless, as before, we define $G^{\boldsymbol P}_\psi$ to be the intersection of all $\mu^{\boldsymbol P}_\psi$-measurable $T \subseteq \preset\omega2$ imaginable from $\boldsymbol P$ with $\mu^{\boldsymbol P}_\psi(T) = 1$.  Let $X = G^{\boldsymbol P}_\psi$.  Then $X$ serves as the canonical test of $\mu^{\boldsymbol P}_\psi$ for $\boldsymbol P$ as applied to $\psi$, and the question `is the sequence of values of the $P_n$s in $X$?', asked of $\psi$, is definitely answered in the affirmative.

The generalization of the foregoing analysis to arbitrary processes of
observation (not necessarily binary) in the quantum theory is straightforward.  We therefore conclude that the apparent stochastic nature of observation, with the squared norm as a probability, is a consequence of the quantum theory with observation treated like any other interaction, \ie, in the many-worlds ``interpretation''.

Interestingly, this aspect of quantum physics sheds light on one of
the singular conceptual difficulties of probability theory, \viz,
that in a process like an infinite sequence of Bernoulli trials,
any specific outcome has probability 0, yet some outcome does
occur.  Similarly, we may conceive of choosing a real number
according to some continuous probability distribution.  Each trial
of this process yields a real number, say $x$, but the probability
of this event, \ie, the measure of the set $\{x\}$, is 0.  Of
course, we have learned to live with this; for example, we
understand that it is not appropriate to define an ``event'' for a
given trial in terms of the result of that trial.  Alternatively,
we may stipulate that we will only consider events that are
definable in some sense; since there are only countably many
definitions, the union of all definable events with probability 0
has probability 0, and we can be reasonably comfortable in the assurance that we will never see a definable event with probability 0 actually occur.

In the quantum theory, however, the situation is a little
different.  Each ``event'' is associated with a selfadjoint
operator, and an event like those just mentioned corresponds to the
0 operator and simply does not happen.  Returning to the scenario presented at the beginning of this article, recall that in answer to the question whether, in a large but finite series of observations, the outcome frequencies might not approximate their expected values, the Copenhagen
interpretation says `yes, but it's unlikely', while the many-worlds
interpretation says `yes, and it certainly happens for some version
of us'.  Let us now consider the question whether, in an
\emph{infinite} series of observations, it is possible for the
outcome frequencies not to tend to their proper limits.  To this
the Copenhagen interpretation says `yes, with probability 0', while
the many-worlds interpretation says `no, absolutely not'.

\section{Absolute continuity of measures}

In Theorem~\ref{eax} we noted that for a $q$-homogeneous $\psi$ the question $P(L_q)$ is answered affirmatively not just for $\psi$ but for any $\psi' \in [\psi]^{\elsalg P}$, the span of the orbit of $\psi$ under the action of $\elsalg P$.  More generally, every $\psi' \in [\psi]^{\elsalg P}$ passes every test of $\mu_\psi$, so that the statistical properties represented by these ``tests'' are really properties of the mode of the system represented by $[\psi]^{\elsalg P}$, not of $\psi$ \emph{per se}.  Of course, in general $\psi'$ may pass additional tests with certainty, \ie, $\mu_{\psi'}$ may have null sets that are not null for $\mu_\psi$.  This section is devoted to an examination of the relation $\psi' \in [\psi]^{\elsalg P}$ in terms of the measures $\mu_{\psi}$ and $\mu_{\psi'}$.  It is not essential to the argument, but is of some interest in its own right.

Suppose $M$ is a nonempty set.  A \emph{$\sigma$-algebra} on $M$ is
a countably complete boolean set-algebra of subsets of $M$
(complementation being understood as relative to $M$).  A
\emph{measure} on a $\sigma$-algebra $\elsalg A$ is a map $\mu : \elsalg A \to [0,\infty]$
such that $\mu(\emptyset) = 0$ and for any $A_0, A_1, \dots \in
\elsalg A$, if $A_m \cap A_n = 0$ for all $m \neq n$, then
\[
\mu \left( \bigcup_{n \in \omega} A_n \right)
 =
\sum_{n \in \omega} \mu(A_n).
\]
Note that we allow $\infty$ as a value of $\mu$, the relevant
arithmetic being that $a + \infty = \infty$ for all $a \in
[0,\infty]$.  $\mu(M)$ is obviously the largest value $\mu$ can
have.

We say that $\mu$ is \emph{$\sigma$-finite} iff $M$ is the union of
countably many sets of finite measure; $\mu$ is \emph{finite}
iff $\mu(M)$ is finite; $\mu$ is a \emph{unit measure}
iff $\mu(M)=1$.  If $\elsalg A$ is the algebra of Borel subsets
of a topological space $\geom X$ (\ie, the closure of the class of open sets under complementation and countable union), we call $\mu$ a \emph{Borel
measure} on $\geom X$.  We are particularly interested in the case
that $M$ is the space $\preset\omega 2$ of functions from $\omega$ to
$2$---\ie, $\omega$-sequences of zeros and ones---and
$\elsalg A$ is the Borel algebra $\elsalg B$ of this space; and we are
most particularly interested in the case of unit Borel measures.

If $M$ and $N$ are sets with respective $\sigma$-algebras $\elsalg M$
and $\elsalg N$ then a map $f : M \to N$ is \emph{$\elsalg M, \elsalg
N$-measurable} iff for any $X \in \elsalg N$, $f \invimage X \in
\elsalg M$.  A function $f : M \to \preset\omega 2$ is \emph{$\elsalg M$-measurable} iff it is $\elsalg
M, \elsalg B$-measurable.

Suppose $\mu$ is a $\sigma$-finite measure on a $\sigma$-algebra
$\elsalg M$ on a set $M$.  We define the integral of a non-negative function as follows.  For any $f : M \to [0,\infty)$, we define
the integral $\int f\, d\mu$ by a simple modification of
the method of Riemann sums familiar from calculus:  instead of
partitioning the domain of $f$ we partition its range, \ie,
$[0,\infty)$, and define $\int f\, d\mu$ as the unique $x \in
\mathbb R$ that lies between the \emph{lower sums}
\[
\sum_{n = 0}^{\infty} x_n \mu( f\invimage [x_n,x_{n+1}] )
\]
and the \emph{upper sums}
\[
\sum_{n = 0}^{\infty} x_{n+1} \mu( f\invimage [x_n,x_{n+1}) ),
\]
for all $0 = x_0 < x_1 < \cdots$ such that $\lim_{n \to \infty} x_n
= \infty$.  Note that the integral so defined may in general be
infinite, although this case will not arise in our applications. 
As a rule, the integral with respect to a measure has all the nice
properties of integration appropriate to this general context.
\begin{definition}
\label{eai}
Suppose $\mu$ and $\nu$ are $\sigma$-finite measures on a
$\sigma$-algebra $\elsalg A$ on a set $M$.  $\nu$ is \emph{absolutely
continuous with respect to $\mu$} iff for all $A \in \elsalg A$,
if $\mu(A) = 0$ then $\nu(A) = 0$; otherwise, $\nu$ is
\emph{singular} with respect to $\mu$. 
\end{definition}
\begin{prop}
\label{eah}
\disptitle{Radon-Nikodym theorem}
Suppose $\mu$ and $\nu$ are $\sigma$-finite measures on a
$\sigma$-algebra $\elsalg M$ on a set $M$.  $\nu$ is absolutely
continuous with respect to $\mu$ iff there exists a non-negative
$\elsalg M$-measurable $F : M \to \mathbb R$ such that for all $X \in
\elsalg M$,\footnote{\normalfont The variable
`$x$' is inserted here to eliminate the ambiguity of expressions
like `$f g$', which might \emph{a priori} refer to either $x
\mapsto f(x) g(x)$, $[ x, y] \mapsto f(x) g(y)$, or
$[ x, y ] \mapsto f(y) g(x)$.}
\[
\nu(X)
 =
\int F(x) \chi^A(x)\, d\mu(x).
\]
\end{prop}
Now suppose $\boldsymbol{P} = \langle P_n \mid n \in \omega
\rangle$ is a sequence of commuting orthogonal projections in a
Hilbert space $\vecsp V$.  Let $P(\cdot)$ be the map from the
set-algebra $\elsalg B$ of Borel subsets of $\preset \omega 2$ to the
complete projection algebra $\elsalg P$ generated by $\langle P_n
\mid n \in \omega \rangle$ defined in Theorem~\ref{cas}.  Note that if $F : \preset\omega2 \to \mathbb R$ is Borel we can
define the integral
\[
\int F(f)\, dP(f)
\]
in the usual way for projection-valued measures, as is done in the
spectral theory of selfadjoint operators, except that $f$ ranges
over $\preset\omega2$ in the present case, rather than over $\mathbb R$.  Note that
\[
P(X)
 =
\int \chi^X\, dP.
\]

For each nonzero $\psi \in \vecsp V$ let $\mu_\psi$ be the
corresponding measure defined by (\ref{eaj}), so
that for any Borel $X \subseteq \preset\omega2$
\[
\| P(X) \psi \|^2 = \mu_\psi(X) \|\psi\|^2.
\]
Suppose $P \in \elsalg P$ and let $\psi' = P \psi$.  Then for any $X
\in \elsalg B$, if $\mu_\psi (X) = 0$, then $P(X) \psi = 0$, so 
\[
P(X) \psi' = P(X) P \psi
 =
P P(X) \psi
 =
0,
\]
so $\mu_{\psi'}(X) = 0$.
\begin{statement}
\label{eal}
In other words, if $\psi' = P \psi$, then $\mu_{\psi'}$ is
absolutely continuous with respect to $\mu_\psi$.
\end{statement}
In general, if $\mathcal S$ is a set of operators on a Hilbert
space $\vecsp V$, a vector $\psi \in \vecsp V$ is said to be
\emph{cyclic for $\mathcal S$ in $\vecsp V$} iff $\vecsp V$ is
the closure of the linear span of the orbit of $\psi$ under the
action of the operators in $\mathcal S$.  In the present case, $\psi$ is cyclic for $\elsalg P$ in $[\psi]^{\elsalg
P}$ but not in any larger subspace of $\vecsp V$.
\begin{thm}
\label{ean}
Suppose $\psi \in \vecsp V$.  For any $\psi' \in [\psi]^{\elsalg P}$, $\mu_{\psi'}$ is
absolutely continuous with respect to $\mu_\psi$.

For any unit measure $\nu$ on $\elsalg B$, if $\nu$ is
absolutely continuous with respect to $\mu_\psi$ then for some
$\psi' \in [\psi]^{\elsalg P}$, $\nu = \mu_{\psi'}$.
\end{thm}
\begin{pf} The first statement is a straightforward generalization of
{\bf\ref{eal}}.  To prove the second assertion suppose $X$ is a Borel subset of $\preset\omega2$.  Let $\chi^X :
\preset\omega2 \to \{0,1\}$ be its characteristic function.  Then, as noted above,
\begin{equation}
\label{eaq}
\int \chi^X\,dP = P(X),
\end{equation}
whence it follows that for any $\psi \in \vecsp V$
\begin{eqnarray}
\left\| \left\{\int \chi^X\, dP \right\} \psi \right\|^2
 &=&
\mu_\psi (X) \|\psi\|^2
 =
\left(\int \chi^X\, d\mu_\psi\right) \|\psi\|^2\\
 &=&
\left(\int \left(\chi^X\right)^2\, d\mu_\psi\right) \|\psi\|^2.
\end{eqnarray}
The last transformation is a trivial one for characteristic
functions, but it gives the proper form to generalize to an
arbitrary non-negative Borel function $G : \preset\omega2 \to
\mathbb R$:
\begin{equation}
\label{ear}
\left\| \left\{ \int G\, dP \right\} \psi \right\|^2
 =
\left(\int G^2\, d\mu_\psi\right) \|\psi\|^2,
\end{equation}
if either side is well defined (in which case they both are).  This
is proved by a straightforward application of limits of sums.

Now suppose $\nu$ is absolutely continuous with respect to
$\mu_\psi$.  Let $F : \preset\omega2 \to \mathbb R$ be a
non-negative Borel function such that for all Borel $X \subseteq
\preset\omega2$\footnote{\normalfont We use
`$f$' as the variable of integration as a reminder that the domain
of integration is not $\mathbb R$ but rather $\preset\omega2$,
which consists of functions from $\omega$ to 2 (= $\{ 0, 1 \}$). 
In this context, however, these are best regarded as points, quite
analogous to real numbers.}
\begin{equation}
\label{eas}
\nu(X)
 =
\int F(f) \chi^X(f)\, d\mu_\psi(f),\end{equation}
as guaranteed by the Radon-Nikodym theorem.  As we have
assumed $\nu$ is a unit measure,
\begin{equation}
\label{eao}
\int F\, dP = 1.
\end{equation}
Note also that $\sqrt F$ is well defined and is non-negative and
Borel.  Let
\begin{equation}
\label{eau}
\psi'
 =
\left\{\int \sqrt F \, dP\right\} \psi.
\end{equation}
Applying (\ref{ear}) and (\ref{eao}) we find that
\begin{equation}
\label{eat}
\|\psi'\|^2
 =
\|\psi\|^2
\end{equation}
(Bounds relating to this identity may be used to show that the
right side of (\ref{eau}) is actually defined.)
Recall that for any Borel $X \subseteq \preset\omega2$,
\[
P(X) = \int \chi^X\, dP.
\]
The following computation is justified by the commutativity of all
the projections involved, and the usual manipulations of integrals
in measure theory (derivable by taking to the limit the
corresponding summation identities).
\begin{eqnarray}
P(X) \psi'
&=&
\left\{\int \chi^X\, dP\right\} \left\{ \int \sqrt F\, dP \right\}
\psi\\
&=&
\left\{ \int \sqrt{F(f)} \chi^X(f)\, dP(f) \right\} \psi
\end{eqnarray}
Taking the squared norms of both sides and applying
(\ref{ear}), noting again that $\left(\chi^X\right)^2 = \chi^X$
since $\chi^X$ is a characteristic function, we obtain
\[
\mu_{\psi'}(X) \|\psi'\|^2
 =
\left( \int F(f) \chi^X(f)\, d\mu_\psi(f) \right) \|\psi\|^2.
\]
We now apply (\ref{eat}) to conclude that
\[
\mu_{\psi'} = \nu,
\]
as desired.\qed\end{pf}

We continue to suppose that $\boldsymbol{P} = \langle P_n \mid n
\in \omega \rangle$ is a sequence of commuting orthogonal
projections in a Hilbert space $\vecsp V$ and that $P(\cdot)$ is
the induced map from the algebra $\elsalg B$ of Borel subsets of
$\preset\omega2$ to orthogonal projections on $\vecsp V$.  Let
$\ker P$ be the kernel or nullspace of $P(\cdot)$, \ie,
the set of Borel sets $X \subseteq \preset\omega2$ for which $P(X)
= 0$.  $\ker P$ is a countably complete ideal in $\elsalg B$.
\begin{statement}
\label{eaw}
Clearly, for any Borel $X \subseteq \preset\omega2$, if $X \in \ker
P$, then $\mu_\psi(X) = 0$ for any $\psi \in \vecsp V$; in other
words, $\ker P$ is included in the null ideal of $\mu_\psi$ for
every nonzero $\psi \in \vecsp V$.  Conversely, if $X \notin \ker
P$, then $\mu_\psi(X) \neq 0$ for some $\psi \in \vecsp V$, and
$\psi$ may be chosen (nonzero) so that $P(X)\psi = \psi$, so that
$\mu_\psi(X) = 1$.
\end{statement}
It is easy to construct examples for which $\ker P$ is exactly the
null ideal of $\mu_\psi$ for some $\psi \in \vecsp V$: simply let
$\vecsp V = [\psi]^{\elsalg P}$.  Clearly these are the
only examples, \ie, the condition $\forall P \in \elsalg P\, (P \ne
0 \implies P\psi \ne 0)$ is equivalent to the condition that $\psi$
be cyclic for $\elsalg P$. 

As far as we know, it is an open question whether this condition is
necessarily satisfied for some $\psi \in \vecsp
V$, but this has little bearing on the issues addressed in this paper.

\section{A critique critiqued}
\label{sec crit crit}

As we noted in the introduction, a version of the preceding analysis was presented by Farhi \emph{et al}\cite{Farhi:1989} in the setting of the tensor product $\vecsp V^{(\infty)} = \vecsp V_0 \otimes \vecsp V_1 \otimes \cdots$ of infinitely many Hilbert spaces.  Their analysis has been criticized by Cassinello and S{\'a}nchez-G{\'o}mez\cite{Cassinello:1996} and Caves and Schack\cite{Caves:2005} as not succeeding in its goal of deriving the probability postulate from prior quantum principles.  In this section we examine this critique, referring specifically to the presentation given in \cite{Caves:2005}.  For the purpose of this commentary, in the interest of brevity, we assume familiarity with \cite{Farhi:1989} and \cite{Caves:2005}.

Following \cite{Farhi:1989}, we restrict our attention to the case that $\vecsp V$ is 2-dimensional.  As noted in the introduction, although $\vecsp V^{(\infty)}$ is a complete inner product space, it is not separable, and is therefore by definition not a Hilbert space.  Indeed, if we suppose for simplicity that $|\psi| = 1$, and we let $\psi_0 = \psi$ and let $\psi_1$ be any normalized vector in $\vecsp V$ orthogonal to $\psi$, then the vectors of the form
\begin{equation}
\label{eq psi infty}
\psi_{j_0} \otimes \psi_{j_1} \otimes \cdots,
\end{equation}
where $\langle j_0, j_1, \dots \rangle$ is an infinite sequence of 0s and 1s, constitute an orthonormal basis for $\vecsp V^{(\infty)}$.  There are as many such sequences as there are real numbers; in particular, there are uncountably many.  Adapting the notations of \cite{Farhi:1989} and \cite{Caves:2005} for ease of reference, we indicate the basis vectors $\psi_0 = \psi$ and $\psi_1$ in $\vecsp V_n$ by $|a,0\rangle$ and $|a,1\rangle$, respectively, and (\ref{eq psi infty}) by $|a;\{j\} \rangle$, $a$ being an arbitrary marker.  (In \cite{Caves:2005} $\psi$ is used in place of $a$.)

$\vecsp V^{(\infty)}$ is canonically the direct sum of separable subspaces, the \emph{components} of $\vecsp V^{(\infty)}$, which are indeed Hilbert spaces.  Let $\vecsp V^{(\infty)}_c$ be the component that contains $|a;\{0,0,\dots\}\rangle = \psi \otimes \psi \otimes \cdots$.  The vectors of the form $|a;\{i\}\rangle$, where $\{i\}$ is a sequence of 0s and 1s that contains only finitely many 1s, constitute an orthonormal basis for $\vecsp V^{(\infty)}_c$.  $\{i\}$ will always be understood to be a sequence with only finitely many 1s.  Note that there are only countably many such sequences.

Let $B$ be the binary observable on $\vecsp V$ in which we are interested, and let $B_n$ be $B$ acting on $\vecsp V_n$ or on the $n^{\mathrm{th}}$ factor of $\vecsp V^{(\infty)}$:
\[
B_n(\psi_0 \otimes \psi_1 \otimes \cdots)
 =
\psi_0 \otimes \cdots \otimes \psi_{n-1} \otimes B \psi_n \otimes \psi_{n+1} \cdots.
\]
In the latter sense $B_n$ corresponds to $P_n$ as used in the preceding sections.  Let $|b,0\rangle$ and $|b,1\rangle$ be normalized 0- and 1-eigenvectors of $B$, respectively.  Let $p_0  = |\langle a,0|b,0\rangle|^2$ and $p_1 = |\langle a,0 | b,1 \rangle|^2 = |\langle \psi | b,1 \rangle|^2 = | B \psi |^2$ (so that $p_1$ is our $q$).  For infinite sequences $\{j\} = \langle j_0, j_1, \dots \rangle$ of 0s and 1s, let $|b;\{j\}\rangle$ be a simultaneous eigenvector of all the $B_n$s with $B_n|b;\{j\}\rangle = j_n |b;\{j\}\rangle$.  Note that $|b;\{j\}\rangle$ is only defined up to a nonzero scalar factor.  As shown in \cite{Farhi:1989}, $|b;\{j\} \rangle$ is not normalizable.

We express the $|a;\{i\}\rangle$s in terms of the $|b;\{j\}\rangle$s by
\begin{equation}
\label{eq a from b}
\langle a;\{i\}| \Psi\rangle
 =
\int d\mu(\{j\})\, \langle a;\{i\} | b; \{j\} \rangle \langle b; \{j\} | \Psi \rangle,
\end{equation}
where $\Psi$ is arbitrary, and the integration is over all sequences $\{j\}$ according to a measure $\mu$, which essentially determines the normalization of $|b;\{j\} \rangle$.  This is (TII) on p. 374 of \cite{Farhi:1989}.

$|b;\{j\}\rangle$ has up to this point only been defined up to a scalar factor, and in \cite{Farhi:1989} this arbitrariness is eliminated (in the parenthesis following (TI)) by setting
\begin{equation}
\label{eq arb one}
\langle a; \{0\} | b; \{j\} \rangle = 1.
\end{equation}
(Here $\{0\}$ is the sequence $\langle 0, 0, \dots \rangle$.  In \cite{Caves:2005}, $| a; \{0\} \rangle$ is called $|\{\psi\} \rangle$.)  With this convention we have (TI) of \cite{Farhi:1989}:
\begin{equation}
\langle a;\{i\} | b; \{j\} \rangle
 =
\prod_{n=0}^\infty
 \frac{\langle a,i_n | b,j_n \rangle}
      {\langle a, 0|b,j_n\rangle},
\end{equation}
and this of course uniquely determines $\mu$, which is shown in \cite{Farhi:1989} to be the $p_1$-homogeneous measure, which we would call $\mu_{p_1}$, \ie, the unit measure that assigns independent weights $p_0, p_1$ to $j_n=0,1$.

\cite{Cassinello:1996} and \cite{Caves:2005} take exception to this argument, noting that the identification of $\mu_{p_1}$ as the measure in (\ref{eq a from b}) depends on the choice (\ref{eq arb one}), which is not mandated by any consideration other than the desire to obtain $\mu_{p_1}$.  \cite{Caves:2005} considers the consequences of omitting this step and defines a function $\{j\} \mapsto c_{\{\psi\}} (\{j\})$ by
\[
\langle a; \{0\} | b;\{j\} \rangle
 =
c_{\{\psi\}}(\{j\}).
\]
The measure $\mu$ of (\ref{eq a from b}) is then
\begin{equation}
\label{eq c psi}
d\mu(\{j\}) = |c_{\{\psi\}}(\{j\})|^2 d\mu_{p_1}(\{j\}),
\end{equation}
which, as pointed out in \cite{Caves:2005}, is rather arbitrary.  We note, however, that this $\mu$ is absolutely continuous with respect to $\mu_{p_1}$, \ie, any set of $\{j\}$s with $\mu_{p_1}$-measure 0 has $\mu$-measure 0.  (We assume that $|c_{\{\psi\}}|^2$ is a measurable function, which is necessary in order that the discussion make sense.)

The error in \cite{Cassinello:1996,Caves:2005} lies in supposing that the argument of \cite{Farhi:1989} depends on $\mu = \mu_{p_1}$.  In fact it does not.  The crux of the argument in \cite{Farhi:1989} comes in the next section, where the frequency operator $F$ corresponding to $B$ is defined by
\[
F|b;\{j\}\rangle = f(\{j\})|b;\{j\}\rangle,
\]
where
\[
f(\{j\})
 =
\frac12\Big(\liminf_{N \to \infty}\frac1N \sum_{n=0}^{N-1} j_n + \limsup_{N \to \infty}\frac1N \sum_{n=0}^{N-1} j_n\Big).
\]
The idea here is that $f(\{j\})$ is defined for all $\{j\}$, and if $\lim_{N \to \infty} \frac1N \sum_{n=0}^{N-1} j_n$ exists then $f(\{j\})$ has that value.

By the strong law of large numbers, for a set of sequences $\{j\}$ with $\mu_{p_1}$-measure 1, the limit does exist and is $p_1$.  It follows that the same is true for any measure absolutely continuous with respect to $\mu_{p_1}$, in particular for the measure $\mu$ defined by (\ref{eq c psi}), which appears in (\ref{eq a from b}) if we do not normalize according to (\ref{eq arb one}).  Hence
\[
F|\Psi\rangle = p_1|\Psi\rangle
\]
for any $\Psi \in \vecsp V^{(\infty)}_c$, in particular for $|a;\{0\}\rangle = \psi \otimes \psi \otimes \cdots$, which is the state of interest.  In other words, a measurement of the proportion of $n$s for which the value of $B_n$ is 1 certainly yields $p_1$.

The same argument applies generally, and all states in $\vecsp V^{(\infty)}_c$ certainly pass all tests of randomness.  Thus the probability postulate does indeed follow from prior quantum principles in this setting, just as it does in the conventional setting as shown in Sections~\ref{sec prob} and \ref{ebk}.

Before moving on, we remark that our discussion at the end of Section~\ref{ebk} is pertinent to Section 4.1 of \cite{Caves:2005}, entitled "Certainty versus probability 1".  The quantum theoretical representation of observables as operators on a (separable) Hilbert space actually \emph{does} make ``certainty'' equivalent to ``probability 1''.  Essentially, quantum observables in the usual sense do not possess sufficient discriminatory power to distinguish these notions.

\appendix\section*{Appendix}

$\emptyset$ is the empty set, \ie, the (unique) set with no
members.  We also use `0' to denote $\emptyset$.  1 is defined as
$\{ 0 \}$, 2 is $\{ 0, 1\}$, 3 is $\{ 0,1,2\}$, etc.  In other
words, the natural numbers are defined in such a way that each
natural number $n$ is the set of numbers that precede it (of which
there are $n$).  These are also referred to as the \emph{finite
ordinals}.  $\omega \eqdef \{ 0, 1, \dots \}$ is the set of all
finite ordinals.

We regard a function as a set of ordered pairs, so that $f = \{ [x,f(x)] \mid x \in \dom f \}$.  $\dom f$, the \emph{domain} of $f$, is the set of first elements of pairs in $f$ and $\im f$, the image of $f$, is the set of second elements of pairs in $f$.  We also represent $f$ by $\langle f(x) \mid x \in \dom f \rangle$.  For $n \in \omega$, a \emph{sequence of length $n$} or \emph{$n$-sequence} or \emph{$n$-tuple} is a function with domain $n$.  In the notation just introduced, if $\sigma$ is an $n$-sequence then $\sigma = \langle \sigma(m) \mid m \in n \rangle$.  We may also denote such a sequence $\sigma$ by indicating its elements as an explicit list between angle brackets.  Thus $\sigma = \langle \sigma(0), \sigma(1), \dots, \sigma(n-1) \rangle$.  Similarly, an \emph{$\omega$-sequence}---also called an \emph{infinite sequence}---is a function with domain $\omega$, and we may indicate such a sequence $\sigma$ by `$\langle \sigma(0), \sigma(1), \dots \rangle$'.  \emph{Concatenation} of finite sequences is defined and indicated as follows:  $\sigma\concat\sigma' = \langle \sigma(0), \dots, \sigma(n-1), \sigma'(0), \dots, \sigma'(n'-1) \rangle$, where $\sigma$ and $\sigma'$ have length $n$ and $n'$, respectively.

If $f$ is a function and $X$ is a set,
\begin{eqnarray*}
f\restriction X &\eqdef& \{ [x, f(x)] \mid x \in X \cap \dom f \}\\
f \image X &\eqdef& \{ y \mid (\exists x \in X \cap \dom f) f(x) = y \}\\
f\invimage X &\eqdef& \{ y \in \dom f \mid f(y) \in X \}.
\end{eqnarray*}  

If $A$ and $B$ are sets we define $\preset{A}B$ (read
``$B$ pre $A$'') to be the set of all functions from $A$ to $B$. 
In particular, for $n \in \omega$, $\preset{n}B$ is the set of functions from $n$ into $B$, \ie, the set of $n$-tuples from $B$, and $\preset{\omega}B$ is the set of all infinite sequences from $B$.  We define $\preset{<\omega}B$ to be $\bigcup_{n \in \omega} \preset nB$.
\begin{definition}
\label{def bool alg}
A boolean algebra is a structure $\elsalg A = ( |\elsalg A|, \compl, \vee, \wedge, 1, 0 )$ with the following properties:
\begin{enumerate}
\item $0 \neq 1$.
\item For all $P, Q, R \in |\elsalg A|$
\begin{eqnarray*}
&P \vee P = P, \quad P \wedge P = P,&\\
&P \vee Q = Q \vee P, \quad P \wedge Q = Q \wedge P,&\\
&P \vee ( Q \vee R ) = (P \vee Q) \vee R,&\\
&P \wedge (Q \wedge R) = (P \wedge Q) \wedge R,&\\
&(P \vee Q) \wedge R = (P \wedge R) \vee (Q \wedge R),&\\
&(P \wedge Q) \vee R = (P \vee R) \wedge (Q \vee R).&
\end{eqnarray*}
\item For each $P \in |\elsalg A|$, $\compl P$ is the unique element of $|\elsalg A|$ such that
\[
P \vee \compl P = 1 \quad\mbox{and}\quad P \wedge \compl P = 0.
\]
\item For all $P, Q \in |\elsalg A|$
\[
\compl(P \vee Q) = \compl P \wedge \compl Q
 \quad\mbox{and}\quad \compl(P \wedge Q) = \compl P \vee \compl Q.
\]
\end{enumerate}
\end{definition}
We write `$P-Q$' for `$P \wedge (\compl Q)$'.

The \emph{order relation} for $\elsalg A$ is defined by
\[
P \leq Q \iff P = P \wedge Q,
\]
or, equivalently, $P \le Q \iff Q = P \vee Q$.  $P \wedge Q$ is therefore the greatest lower bound and $P \vee Q$ the least upper bound of $\{ P, Q \}$. 

A boolean algebra $\elsalg A$ is
\emph{complete} iff if every subset of $\elsalg A$ has a greatest
lower bound, which we call the \emph{meet} of the set.  Every set
in a complete boolean algebra also has a least upper bound, which
we call the \emph{join} of the set.  $\elsalg A$ is \emph{countably complete} iff every countable set of elements has a meet (equivalently, a join) in $\elsalg A$.

Elements of a boolean algebra $\elsalg A$ are \emph{compatible} iff their meet is nonzero.  $\elsalg A$ has the \emph{countable chain condition} iff every set of pairwise incompatible elements is countable.
\begin{prop}
\label{dun}
If a boolean algebra $\elsalg A$ is countably complete and satisfies the
countable chain condition then $\elsalg A$ is complete.  
\end{prop}
\begin{pf} The proof depends on the axiom of choice.  On this
assumption any set can be put in one-one correspondence with a
von Neumann cardinal, \ie, an ordinal $\kappa$ such that there does not exist a function from any ordinal $\lambda < \kappa$ onto $\kappa$.  We proceed by induction.  Suppose joins
exist for all sets of cardinality less than $\kappa$, where
$\kappa$ is a cardinal, and suppose $X = \{ x_\alpha \mid \alpha
\in \kappa \}$ is a subset of $\elsalg A$.  Let $y_\alpha =
\bigvee \{ x_\beta \mid \beta \in \alpha \}$ for each $\alpha \in
\kappa$.  The elements $x_\alpha - y_\alpha$, $\alpha \in
\kappa$, are pairwise incompatible, so by the countable chain
condition, only countably many of them are nonzero, so the join
of the corresponding $y_\alpha$s exists.  This is clearly the
least upper bound of $X$.\qed\end{pf}


\begin{thebibliography}{1}

\bibitem{Cassinello:1996}
A.~Cassinello and J.~L. S{\'a}nchez-G{\'o}mez.
\newblock On the probabilistic postulate of quantum mechanics.
\newblock {\em Foundations of Physics}, 26:1357--1374, 1996.

\bibitem{Caves:2005}
C.~M. Caves and R.~Schack.
\newblock Properties of the frequency operator do not imply the quantum
  probability postulate.
\newblock {\em Annals of Physics}, 315:123--146, 2005.

\bibitem{Everett:1957}
H.~Everett, III.
\newblock {"Relative State"} formulations of quantum mechanics.
\newblock {\em Rev Mod Phys}, 29:454, 1957.

\bibitem{Farhi:1989}
E.~Farhi, J.~Goldstone, and S.~Gutmann.
\newblock How probability arises in quantum mechanics.
\newblock {\em Annals of Physics}, 192:368--382, 1989.

\bibitem{Graham:1973}
N.~Graham.
\newblock {\em The Many Worlds Interpretation of Quantum Mechanics}, page 229.
\newblock Princeton University Press, Princeton, 1973.

\bibitem{Gutmann:1995}
S.~Gutmann.
\newblock Using classical probability to guarantee properties of infinite
  quantum sequences.
\newblock {\em Physical Review A}, 52(5):3560--3562, 1995.

\bibitem{Hartle:1968}
J.~B. Hartle.
\newblock Quantum mechanics of individual systems.
\newblock {\em Am. J. Phys.}, 36:704, 1968.

\bibitem{RVWhv:2005}
R.~A. Van~Wesep.
\newblock Hidden variables in quantum mechanics: Generic models, set-theoretic
  forcing, and the emergence of probability.
\newblock arXiv reference quant-ph/0506040.

\bibitem{vonNeumann:1938}
J.~von Neumann.
\newblock On infinite direct products.
\newblock {\em Composition Mathematica}, 6:1, 1938.

\end{thebibliography}

\end{document}